\documentclass[baaa]{baaa}

\usepackage[pdftex]{hyperref}
\usepackage{subfigure}
\usepackage{natbib}
\usepackage{helvet,soul}
\usepackage[font=small]{caption}

\contriblanguage{1}

\contribtype{1}

\thematicarea{7}

\received{\ldots}
\accepted{\ldots}

\title{The rise and fall of radio-luminous merger shocks\\ from a large sample of galaxy cluster simulations}

\titlerunning{Radio shock evolution in merging galaxy clusters}

\author{
S.E.~Nuza\inst{1,2}
}

\authorrunning{S.E.~Nuza}

\contact{snuza@iafe.uba.ar}

\institute{
Instituto de Astronom\'ia y F\'isica del Espacio, CONICET--UBA, Argentina
\and
Consejo Nacional de Investigaciones Cient\'ificas y T\'ecnicas, Argentina
}

\resumen{La generación de ondas de choque es una consecuencia natural del proceso jerárquico de formación de estructura. Con el tiempo, las estructuras cósmicas crecen en masa y tamaño mediante fusiones y a través de la acreción continua de material en los pozos de potencial de los halos de materia oscura. En particular, algunos cúmulos galácticos dinámicamente perturbados exhiben radio estructuras no térmicas conocidas como {\it radio relics}, que se cree están ligadas a choques de fusi\'on en diferentes etapas de evolución. Estas ondas de choque brillan como resultado de la aceleración de electrones relativistas en presencia de campos magnéticos. En esta contribución, se analiza una muestra de re-simulaciones hidrodinámicas de cúmulos galácticos en un contexto cosmológico pertenecientes al proyecto {\sc The Three Hundred Project} con el fin de estudiar la evolución de los {\it relics} en función de la masa cumular y el corrimiento al rojo. Esta muestra sintética nos permite calcular en detalle la radio luminosidad de los choques durante fusiones de cúmulos galácticos desde su generación al comienzo de la fusión hasta su desaparición.}

\abstract{The generation of merger shocks is a natural outcome of the hierarchical process of structure formation. As time elapses cosmic structures grow in mass and size via mergers and through the continuous accretion of material onto the potential wells of dark matter haloes. In particular, some dynamically-perturbed galaxy clusters exhibit spectacular non-thermal radio features known as radio relics that are believed to trace cluster merger shocks at different stages of evolution. These radio shocks are thought to be illuminated by the acceleration of cosmic ray electrons in the presence of intracluster magnetic fields. In this contribution, we analyse a large sample of hydrodynamical, cosmological re-simulations of merging galaxy clusters belonging to {\sc The Three Hundred Project} to study the median evolution of radio relics as a function of cluster mass and redshift. This synthetic cluster merger sample enables us to compute in detail the luminosity output of radio shocks from their onset at core passage to demise.}


\begin{document}

\maketitle
\section{Introduction}
\label{intro}

\begin{figure*}[!t]
\hspace{1cm}
%\centering
\includegraphics[width=0.8\textwidth]{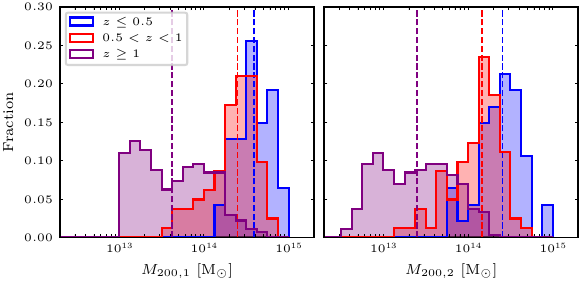}
\caption{\emph{Left-hand panel:} Mass distribution of the most massive progenitor, $M_{200,1}$, in our galaxy cluster merger sample for 3 different redshift bins. \emph{Right-hand panel:} Idem for the secondary progenitor mass, $M_{200,2}$. Dashed vertical lines in both panels indicate the median mass value of each histogram.}
\label{fig1}
\end{figure*}

In a hierarchical formation scenario of cosmic structures large dark matter haloes grow as a result of accretion and mergers with other systems. In this context, shock waves are naturally produced during halo assembly in a wide range of scales. In particular, galaxy clusters are believed to experience a series of major mergers during their lifetime, leading to the formation of merger shock waves which are ideal sites for (re)acceleration of cosmic ray electrons (CRe). In the presence of intracluster magnetic fields, these CRe produce non-thermal radio emission that trace large-scale shock waves in the outskirts of galaxy clusters. These so-called radio relics are preferentially observed in dynamically-perturbed systems showing clear signs of interaction \citep[e.g.,][]{Nuza12}.   

Previous studies on the modelling of radio relic populations from samples of simulated galaxy clusters according to the current formation scenario \cite[e.g.,][]{Hoeft11} focused, for instance, on their occurrence as a function of cluster mass and redshift \citep{Nuza12}, scaling relations \citep{Nuza12,Nuza17,Lee24}, polarization structure \citep{Wittor19} and spectral properties \citep{Wittor21}. These radio relic samples are typically built for a limited number of snapshots drawn from large cosmological simulation volumes, thus limiting the study of the evolution of radio relic properties at high-temporal resolution. 

Recently, \cite{Nuza23} benefited from the larger temporal sampling of a galaxy cluster re-simulation belonging to the so-called {\sc Three Hundred Project} \citep{Cui18} to build radio relic light curves, computing, for the first time, the expected output of non-thermal radio emission obtained from galaxy cluster mergers during the complete simulated cosmic evolution. A follow-up paper by \cite{Nuza24} expanded this study to a sample of several hundred cluster regions estimating average trends for a large sample of groups and clusters focusing on radio luminosity light curves and radio relic power and cluster mass relations. 

In this contribution, we further explore on several aspects concerning the behaviour of average radio relic light curves during the approach and merging phase of colliding systems characterised by two main structures extracted from the galaxy cluster sample of \cite{Nuza24}, focusing on the time evolution of the expected radio power. In particular, we concentrate on cluster mass, redshift distribution and initial configuration of the sample and the radio luminosity output from merger shocks in galaxy groups and clusters according to our modelling of non-thermal radiation.  

This proceeding is organised as follows. In Sect.~\ref{simu} we discuss the galaxy cluster merger sample used in this work. In Sect.~\ref{radio_model} we briefly describe the non-thermal radio emission model and the shock finder in the simulations. In Sect.~\ref{results} we discuss our results. Finally, in Sect.~\ref{concl} we present the conclusions.

\section{Galaxy cluster merger catalogue}
\label{simu}

In this work, we used a sample of galaxy cluster regions that belongs to the {\sc Three Hundred Project}, a suite of 324 spherical zoom-in re-simulations of galaxy cluster regions of radius $15\,h^{-1}\,$Mpc extracted from the $1\,h^{-3}\,$Gpc$^3$ MDPL2
MultiDark simulation \citep{Klypin16}, where $h$ is the reduced Hubble constant. The cosmological model adopted is consistent with the Planck 2015 cosmology \citep{Planck16}. Galaxy cluster regions were re-simulated with the {\sc Gadget-X} code \citep[e.g.,][]{Beck16} including full hydrodynamics, metal-dependent gas cooling, an UV background radiation field, and other relevant sub-grid astrophysical processes such as star formation, feedback from supernovae/active galactic nuclei and black hole growth. Mass resolution at the beginning of each re-simulation is $1.27\times10^9\,h^{-1}\,$M$_{\odot}$ and $2.36\times10^8\,h^{-1}\,$M$_{\odot}$, for dark matter and gas particles, respectively.  

From the simulated cluster regions, we built a catalogue of mergers displaying a significant mass increase of $\Delta M/M\geq0.5$ within a time interval equal to the cluster dynamical time or less \citep[][]{Contreras22}. Despite the great variety of astrophysical situations possible during galaxy cluster mergers (sometimes involving more than two objects), here we simply characterise mergers by accounting for the two main merging systems. We note that the particular mass fraction increase chosen guarantees that our merger sample mainly corresponds to major mergers, i.e. $\mu\equiv M_{200,2}/M_{200,1}\gtrsim0.3$, with $M_{200,1}$ and $M_{200,2}$ the mass of the main and secondary clusters, respectively. As a result, we end up with a sample of 555 well-resolved mergers with $M_{200,1}\geq 10^{13}\,$M$_{\odot}$ within $R_{200,1}$, i.e. the main progenitor's radius enclosing an overdensity of 200 times the critical density of the Universe.    

\subsection{Cluster merger sub-samples}
\label{merger_samples}

Additionally, we split the merger catalogue into different sub-samples. In this way, we are able to study the impact of cluster mass and redshift of the merging systems in the final radio light curves. First, we select i) merging groups for systems satisfying  $10^{13} \leq M_{200,1} / \mathrm{M}_{\odot} < 10^{14}$ and ii) merging clusters with $M_{200,1} \geq 10^{14}\,\mathrm{M}_{\odot}$. Given that most of the collisions in our catalogue correspond to major mergers, this choice ensures that main and secondary masses are of the same order, thus naturally separating the sample into less/more massive mergers. Second, we also built additional sub-samples selecting systems by merger redshift instead of cluster mass. The redshift of each merger in our catalogue is computed as the mean value between the redshifts corresponding to $t_{\rm start}$ and $t_{\rm end}$, the start and end times of each merger \citep[see the discussion in][]{Contreras22}. Specifically, we also take iii) early and iv) late merger samples by imposing the conditions $z\geq z'$ and $z<z'$, respectively. To assess the evolution of merging systems across the whole cosmic history, we simply take $z'=1$ as most of the less massive clusters in our sample lie above this value (see Fig.~\ref{fig1}). 

\section{The radio output from shocks}
\label{radio_model}

The radio emission produced in structure formation shocks is computed following \cite{Nuza17}. In this model, a fixed fraction of thermal electrons get accelerated at the shock fronts acquiring an energy distribution consistent with the diffusive shock acceleration (DSA) scenario. The resulting CRe are then advected by the downstream plasma, where they experience radiative losses, primarily through synchrotron radiation and inverse Compton scattering with cosmic microwave background (CMB) photons. The non-thermal emission behind the shock is obtained by integrating the electron distribution along the
downstream region assuming a constant shock velocity,
which can be related to the sonic Mach number, $\mathcal{M}$. Then, the radio power, $P_{\nu,A}$, per unit frequency, $\nu$, and area, $A$, for each gas particle scales as

\begin{equation}
    P_{\nu,A} \propto n_{\rm e}\,\nu^{-\frac{s}{2}}\,T_{\rm d}^{\frac{3}{2}}B_{\rm d}^{1+\frac{s}{2}}\left(B_{\rm CMB}^2+B_{\rm d}^2\right)^{-1}\,\Psi({\cal M}){\rm ,}
\end{equation}

\begin{figure}[!t]
%\centering
\hspace{-0.1cm}
\includegraphics[width=0.95\columnwidth]{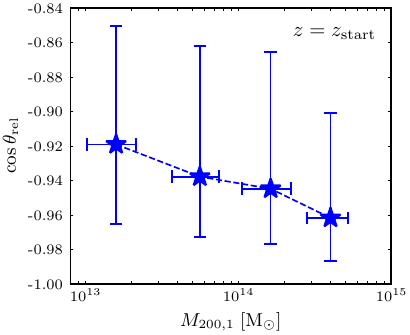}
\caption{Median value of the cosine of the angle between the relative position and velocity vectors of the two main progenitors in the cluster sample as a function of main progenitor mass computed at the starting redshift of the merger (see Sect.~\ref{mass_orbit}). Horizontal error bars indicate the standard deviation in each bin, whereas vertical error bars stand for the $25$th and $75$th percentiles.}
\label{fig2}
\end{figure}

\noindent where $n_{\rm e}$ is the electron density, $s$ is the slope of electron energy distribution given by DSA, $T_{\rm d}$ is the post-shock temperature, $B_{\rm d}$ is the post-shock magnetic field, $B_{\rm CMB}$ is the magnetic measure of the CMB energy density, and $\Psi(\cal M)$ depends on the shock strength strongly suppressing radio emission for $\mathcal{M}\lesssim2$. 

Magnetic fields are assumed to scale with local electron density following a profile consistent with observations of the Coma cluster \citep{Dolag01,Bonafede10}. Shock fronts in the simulations are identified from gas particles fulfilling a set of criteria such as convergent flows and entropy/density jumps and the Mach number is computed using the Rankine-Hugoniot conditions \citep{Landau59}. Further details on the non-thermal radio model and the shock finder can be found in \cite{Nuza17} and references therein.

\section{Results}
\label{results}

\subsection{Cluster mass distribution and merger orbits}
\label{mass_orbit}

Figure~\ref{fig1} shows the distribution of the progenitor masses during galaxy cluster mergers in our sample for 3 different redshift bins for the complete merger catalogue. In particular, the merger ratio, $\mu'=\tilde{M}_{200,2}/\tilde{M}_{200,1}$, of the median cluster mass values, for all redshift intervals is $\mu'\gtrsim0.6$ demonstrating that many of the two main merging structures in our cluster merger sample are similar. Additionally, there is a clear trend of progenitor masses with redshift, where the less (more) massive systems are located at larger (smaller) redshifts. Specifically, the most massive galaxy clusters in the sample are formed at redshifts $z\leq0.5$, reaching masses similar to that of massive local clusters in agreement with observations. This is a natural consequence of the hierarchical nature of structure formation where the most massive systems form as a result of the coalescence of less massive objects.

The normalised projection of the relative velocity of the two main merger progenitors onto the direction of the radial vector between them, $\cos\theta_{\rm rel}$, is shown in Fig.~\ref{fig2} as a function of main progenitor mass. Here, $\theta_{\rm rel}$ is the relative angle between the two vectors computed at the start time of the cluster merger according to the time definition of \cite{Contreras22}. This quantity provides an indication on the nature of the orbit followed by the two merging structures with $\cos\theta_{\rm rel}=-1$ corresponding to a purely radial orbit. As seen in Fig.~\ref{fig2}, the vast majority of systems in our catalogue, from galaxy groups to clusters, show fairly radial orbits at the beginning of the interaction, which translates into typical impact parameters of about a few hundred kpc at core passage. This is expected as colliding structures within a cosmological context usually come from preferred directions determined by large-scale cosmic filaments connected to the central clusters. This trend, however, is slightly less pronounced for the less massive clusters, which are more likely to experience perturbations at an early stage of the merger phase.  

\begin{figure}[!t]
%\centering
\hspace{-0.4cm}
\includegraphics[width=0.99\columnwidth]{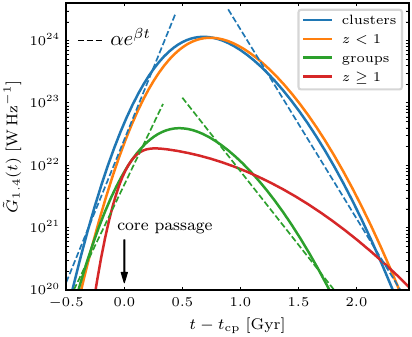}
\caption{Median radio light curve evolution at $1.4\,$GHz during galaxy cluster collisions after subtraction of the linear baseline (see Sect.~\ref{lcs}) as a function of time with respect to the core passage for the merger sub-samples introduced in Sect.~\ref{merger_samples}: i) group ($10^{13} \leq M_{200,1}/\mathrm{M}_{\odot} < 10^{14}$); ii) cluster ($M_{200,1}\geq 10^{14}\,$M$_{\odot}$); iii) early ($z\geq1$), and iv) late ($z<1$) mergers. Dashed lines correspond to best-fitting exponential functions of the cluster and group sub-samples for times before and after the peak of each light curve.}
\label{fig3}
\end{figure}

\subsection{Radio luminosity evolution from an ensemble of galaxy cluster mergers}
\label{lcs}

To study the average radio luminosity output produced during cluster mergers, we followed previous works and computed all radio emission at $1.4\,$GHz produced by shocks in our galaxy cluster catalogue within a distance of $R_{200,1}(t)$ from the main cluster's centre according to the non-thermal model introduced in Sect.~\ref{radio_model} during the complete merger evolution. This distance is adequate to encompass the temporal evolution of merger radio shocks, from their onset to demise, without including too much radiation produced in external shocks. As a result, we obtained a sample of radio light curves that can be combined taking into account the core-passage time, $t_{\rm cp}$, of each system. This parameter was estimated assessing the relative distance and velocity of the two most massive progenitors during the collisions. In this way, it is possible to estimate the median radio luminosity light curve of cluster mergers as a function of cluster mass and redshift. 

The median radio relic luminosity during cluster mergers from simulation data is described with the following expression:

\begin{equation}
    \tilde{P}_{1.4}(t)=\tilde{G}_{1.4}(t)+\tilde{l}_{1.4}(t){\rm ,}
\end{equation}

\noindent where $\tilde{G}_{1.4}(t)$ is a skewed Gaussian function chosen to reproduce the radio luminosity evolution and $\tilde{l}_{1.4}(t)$ is a linear baseline to mimic the cumulative effect of spurious radiation (i.e., not related to the merger shocks) that may be produced within the cluster volume.

Figure~\ref{fig3} shows the median radio light curve as a function of time with respect to the core passage after subtraction of the linear function, $\tilde{l}_{1.4}(t)$, for the 4 merger sub-samples defined in Sect.~\ref{merger_samples}. In general, a strong increase of the radio luminosity output can be seen for all curves after core passage in a period of about several hundred Myr. This is highlighted by the exponential fits shown as dashed lines during the radio growth phase. In particular, the most massive mergers belonging to the cluster sub-sample (blue solid line) can boost the radio shock emission up to about $\sim2$ orders of magnitude during the growth phase. Similarly, the late sub-sample (orange solid line) shows a similar evolution as most of the massive systems are typically assembled at lower redshifts. After the emission peak, in all cases, radio emission decays in somewhat longer timescales as a result of skewness of the light curves. Smaller mass mergers are represented by the group (green solid line) and early (red solid line) sub-samples which present a similar evolution, albeit reaching much lower radio luminosity peaks. Although the radio growth phase in these two sub-samples is comparable, the decay phase after the peak is less pronounced in the early sub-sample, most likely owing to the inclusion of highly-perturbed, lower-mass systems at high redshift. In total, radio shocks in our sub-samples require $\sim1.2–1.6\,$Gyr to recover luminosities matching core passage levels.

\section{Conclusions}
\label{concl}

In this work, we made use of the large temporal sampling of an ensemble of galaxy cluster re-simulations from {\sc The Three Hundred Project} to study the radio luminosity output of merger shocks during galaxy cluster collisions. Our catalogue comprises a total of 555 major mergers within a cosmological context primarily characterised by the two main colliding structures, although more complicated situations are not uncommon.  In general, most of the massive mergers take place at lower redshifts in agreement with the hirarchical nature of the accepted structure formation scenario. The colliding systems start the merger phase in almost radial orbits with a small tangential velocity component that translates into impact parameters of about a few hundred kpc at core passage. 

After splitting the merger catalogue into several sub-samples comprising less and more massive systems, we computed the median radio light curves of relics during galaxy cluster mergers to characterise the radio luminosity output from the onset of merger shocks, just before core passage, to demise after $\gtrsim1\,$Gyr, moderately depending on cluster mass. We found that typical light curves rise abruptly after core passage until the emission peak is reached $\sim500-700\,$Myr after core passage. Overall, merger shocks represent the tip of the iceberg of the non-thermal radio Universe, primarily shaped by the highest galaxy density regions.

%%%%%%%%%%%%%%%%%%%%%%%%%%%%%%%%%%%%%%%%%%%%%%%%%%%%%%%%%%%%%%%%%%%%%%%%%%%%%%
% Para figuras de dos columnas use \begin{figure*} ... \end{figure*}         %
%%%%%%%%%%%%%%%%%%%%%%%%%%%%%%%%%%%%%%%%%%%%%%%%%%%%%%%%%%%%%%%%%%%%%%%%%%%%%%

\begin{acknowledgement}
S.E.N. is a member of the Carrera del Investigador Cient\'{\i}fico of CONICET. He acknowledges support from CONICET (PIBAA R73734), Agencia Nacional de Promoci\'on Cient\'{\i}fica y Tecnol\'ogica (PICT 2021-GRF-TI-00290) and UBACyT (20020170100129BA).
\end{acknowledgement}

%%%%%%%%%%%%%%%%%%%%%%%%%%%%%%%%%%%%%%%%%%%%%%%%%%%%%%%%%%%%%%%%%%%%%%%%%%%%%%
%  ******************* Bibliografía / Bibliography ************************  %
%                                                                            %
%  -Ver en la sección 3 "Bibliografía" para mas información.                 %
%  -Debe usarse BIBTEX.                                                      %
%  -NO MODIFIQUE las líneas de la bibliografía, salvo el nombre del archivo  %
%   BIBTEX con la lista de citas (sin la extensión .BIB).                    %
%                                                                            %
%  -BIBTEX must be used.                                                     %
%  -Please DO NOT modify the following lines, except the name of the BIBTEX  %
%  file (whithout the .BIB extension).                                       %
%%%%%%%%%%%%%%%%%%%%%%%%%%%%%%%%%%%%%%%%%%%%%%%%%%%%%%%%%%%%%%%%%%%%%%%%%%%%%% 

\bibliographystyle{baaa}
\small
\bibliography{bibliografia}
 
\end{document}